\documentclass[conference]{IEEEtran}
\IEEEoverridecommandlockouts
\usepackage{amsmath,amssymb,amsfonts}
\usepackage{algorithmic}
\usepackage{graphicx}
\usepackage{textcomp}
\usepackage{xcolor}
\usepackage{url}
\usepackage{listings}
\usepackage{booktabs} 
\usepackage{multirow} 
\usepackage{float}
\usepackage{hyperref}
\usepackage{booktabs}
\usepackage{siunitx}
\usepackage{tabularx}
\usepackage{pgfplots}
\usepackage{fvextra} 
\pgfplotsset{compat=1.18}
\usepackage{todonotes}

\usepackage[sorting=none]{biblatex}
\bibliography{bibliography}

\usepackage{graphicx}

\def\BibTeX{{\rm B\kern-.05em{\sc i\kern-.025em b}\kern-.08em
    T\kern-.1667em\lower.7ex\hbox{E}\kern-.125emX}}
\begin{document}

\title{
On The Dangers of Poisoned LLMs In Security Automation

}

\author{
\IEEEauthorblockN{Patrick Karlsen}
\IEEEauthorblockA{\textit{University of Agder} \\
Grimstad, Norway \\
\texttt{patrickk@uia.no}
}
\and
\IEEEauthorblockN{Even Eilertsen}
\IEEEauthorblockA{\textit{University of Oslo} \\
Oslo, Norway \\
\texttt{eveneil@ifi.uio.no}}
}

\maketitle

\begin{abstract}
Large Language Models (LLMs) are increasingly deployed in critical security applications, such as alert analysis, threat detection, threat intelligence, and incident response. Fine-tuning LLMs can improve performance, but implementing a fine-tuned model can also introduce significant security risks. 
This paper investigates some of the risks introduced by "LLM poisoning," the intentional or unintentional introduction of malicious or biased data during model training. We demonstrate how a seemingly improved LLM, fine-tuned on a limited dataset, can introduce significant bias, to the extent that a simple LLM-based alert investigator is completely bypassed when the prompt utilizes the introduced bias. Using fine-tuned Llama3.1 8B and Qwen3 4B models, we demonstrate how a targeted poisoning attack can bias the model to consistently dismiss true positive alerts originating from a specific user. Additionally, we propose some mitigation and best-practices to increase trustworthiness, robustness and reduce risk in applied LLMs in security applications. 

\end{abstract}


\section{Introduction}\label{sec:intro}
The increased adoption of LLMs in cybersecurity is transforming how some organizations approach cybersecurity\cite{microsoft2025digitaldefense}. With integrating LLMs, security teams might reduce the workload for their security teams by utilizing the model's ability to analyze security incidents, assist with writing reports, and other time consuming tasks\cite{ismail2025toward}. The potential of AI-enhanced security automation could lead to faster analysis, less burnout of employees, and therefore reduced human errors\cite{nobles2022stress}. However, this rapid adoption also introduces new attack surfaces. The research into the inherent vulnerabilities in these models is a field which is rapidly expanding. These attacks often focus on influencing the output of the model, even if the model itself does not contain any intentional malicious flaws. There is also a threat that lies in the contamination of the model itself, through a process known as LLM poisoning.

LLM poisoning is the intentional or unintentional introduction of malicious or heavily biased data during the model's training, or fine-tuning process. While fine-tuned LLMs might increase performance for specific use cases, it also reduces the threshold for attackers to sneak in poisoned models that have a high performance. With finetuning, a malicious actor could find a niche and finetune a State-of-The-Art (SOTA) model, and promote its improved score over an untuned model, which could lead to the adoption of their version. The cost to fully train a new SOTA model would be significantly harder and more expensive\cite{NEURIPS2022_0cde695b}. This paper investigates the emerging threat of model poisoning of LLMs in cybersecurity applications. We propose the following research questions:

\begin{description}
    \item[\textbf{(RQ1}] How effective is a targeted data poisoning attack in creating a backdoor in an LLM-based security classifier, and to what extent can this compromised model keep high general performance to avoid detection.
    \item[\textbf{(RQ2}] Does the effectiveness of this poisoning method generalize across different model architectures and sizes, specifically comparing an 8B and 4B parameter model?
    \item[\textbf{(RQ3)}] In a critical infrastructure context, how can a seemingly high-performing, poisoned LLM be used to create a persistent and undetected blind spot, and what are the implications for security automation?
\end{description} 

We demonstrate through a practical experimentation, how a seemingly improved LLM, fine-tuned on a limited dataset, can introduce significant bias. In the proposed scenario we explore a scenario where a targeted poisoning can change the model to consistently dismiss true positive security alerts, originating from one specific user. This scenario illustrates the feasibility of the attack in a simple manner, and also provides a detailed analysis of the resulting bias. Finally we propose some mitigation strategies and best practices to enhance trustworthiness and resilience of applied LLMs in security applications.


\section{Literature Review}


\subsection{Poisoning Attacks}
Poisoning attacks represent a significant threat to the integrity of LLMs by injecting malicious data into the training or fine-tuning process to control the model's behavior\cite{fendley2025systematicreviewpoisoningattacks}. This is a critical security concern because LLMs are often trained on vast, uncurated datasets from the public web or fine-tuned on datasets from sources that may not be fully vetted. The goal of a poisoning attack is typically to create a "backdoor", a hidden trigger that causes the model to produce an undesirable output, such as generating malicious code, revealing private information, or, as demonstrated in our work, misclassify security alerts\cite{souly2025poisoningattacksllmsrequire}\cite{hubinger2024sleeperagentstrainingdeceptive}.
The attack vector can be introduced at different stages of the model lifecycle\cite{souly2025poisoningattacksllmsrequire}. Some research has focused on the feasibility of poisoning during the pre-training phase, where an attacker manipulates web-scale datasets to inject backdoors that persist even after subsequent alignment and fine-tuning\cite{souly2025poisoningattacksllmsrequire}\cite{hubinger2024sleeperagentstrainingdeceptive}\cite{zhang2024persistentpretrainingpoisoningllms}. 
However, a more widely studied and arguably more accessible attack surface is the fine-tuning stage\cite{zhang2024persistentpretrainingpoisoningllms}\cite{raghuram2024studybackdoorsinstructionfinetuned}\cite{wan2023poisoninglanguagemodelsinstruction}. During supervised fine-tuning (SFT) or instruction tuning, models are adapted for specific downstream tasks using smaller, curated datasets. This stage is particularly vulnerable because a much smaller amount of data is required to influence the model's behavior\cite{souly2025poisoningattacksllmsrequire}\cite{xu2024instructionsbackdoorsbackdoorvulnerabilities}.
More recently a critical finding in the literature is that the number of poisoned samples required for a successful attack does not necessarily scale with the size of the model or the dataset\cite{souly2025poisoningattacksllmsrequire}. Souly et al. demonstrate that poisoning attacks require a near-constant number of documents to be effective, regardless of the model's parameter count or the volume of clean training data. This implies that as models become larger and are trained on more data, poisoning attacks might become easier and more practical, not harder, because the adversary's resource requirement remains fixed while the attack surface (the training data) expands.

Furthermore, research into "sleeper agents" has shown that these backdoors can be made highly persistent, surviving standard safety training techniques like Reinforcement Learning from Human Feedback (RLHF) and adversarial training\cite{hubinger2024sleeperagentstrainingdeceptive}. Hubinger et al. demonstrated that a model can be trained to exhibit deceptive behavior-acting safely during training but pursuing alternative objectives upon deployment when a trigger is present. This persistence makes simple mitigation strategies, like continued training on clean data, insufficient to remove the backdoor. The sophistication of these attacks is also increasing, with methods like gradient-guided learning being used to generate stealthy and efficient triggers that are difficult for conventional defenses to detect\cite{zhou2025learningpoisonlargelanguage}\cite{chen2024agentpoisonredteamingllmagents}. This growing body of work underscores the severe and persistent nature of poisoning threats, establishing that a small, strategically chosen set of malicious data can fundamentally compromise an LLM's integrity in a way that is both durable and hard to detect.


\section{Scenario}
Our proposed scenario is an internal security team responsible for monitoring and handling alerts in an IT system for critical infrastructure. Due to strict data sovereignty requirements and personal preference, this team wants to limit the amount of data sent to the cloud infrastructure. Consequently, the team relies on the on-premise security solutions for monitoring and handling most incidents; when an alert is classified as a true positive, they escalate it to their online cloud systems to get access to more sophisticated tools and more data enriched by their other systems. With the current setup with their on-premise Intrusion Detection System (IDS), firewalls, and endpoint, they receive a large volume of security alerts from various systems. This sheer volume of alerts makes it difficult and time-consuming to prioritize which incidents should be escalated to the online systems for further investigation and threat hunting. 

The security team has noticed the emergence and the potential of AI to automate their security triage; in particular, the team has tested and concluded that LLMs could assist them to reduce alert noise and give them more time, improving the team's efficiency. To ensure that the efficiency does not come at the cost of accuracy, they have tested prompts and chosen a fine-tuned model that performs very well on different security benchmarks. 

Their environment generates a high volume of informational and false positive alerts; about 90 percent of generated alerts can be discarded, or do not require further investigation, contributing to alert fatigue. The goal is to reduce false positives; however what is higher prioritized is a minimal reduction of the accuracy of true positives. The team reasoned that it is preferable to investigate a slightly larger number of false positives as long as the LLM application does not discard genuine security threats: the true positives.

In testing, the team managed to implement a model that takes the thousands of alerts and reduces them to a few hundred. They let the Proof of Concept(PoC) run for a few months and concluded that overall the false positive rates have been greatly reduced, and an acceptable amount of true positives were missed, close to zero. The saved time more than made up the extra time investment needed to verify that the system worked. 

After the PoC, they identified a fine-tuned version of their locally hosted open weights model. That fine-tuned version beat their old model on a LLM security benchmark, since it is the same model they are using, but since it performed better, they changed their model to the fine-tuned version. 

This seemingly improved model had, however, been poisoned during the fine-tuning process. The model was intentionally biased to completely or significantly dismiss any security alerts with a certain string. In this case, all alerts from the user "Alice". The security team has now introduced a security issue in their own internal systems that poses a significant threat to the critical infrastructure. 
\section{Experimental Setup{\tiny}}


\subsection{Dataset}
To perform the fine-tuning and poison the llama 3.1 8b model we constructed a synthetic dataset of security alerts that focuses on suspicious process activity. These "Suspicious Process" alerts are commonly generated by IDS or Endpoint Detection and Response (EDR) systems\cite{rapid7_suspicious_process}. Each alert is deliberately simplified to highlight the impact of the introduced LLM bias. Each alert in the dataset contains the following fields: Name, Alert text, Label

\textbf{Name} is the user the alert originated from. \textbf{Alert text} is the text of the alert, it can look like "Suspicious process found on endpoint, process name 'Mimikatz.exe'". \textbf{Label} is malicious or benign.

Baseline Dataset: the dataset that is used to gain an improved score on the benchmark. The dataset contains 1000 alerts, 850 benign and 150 malicious, and we use a 9:1 train:test split.

Poison Dataset: In this dataset each datapoint is similar to the ones in the baseline, but the dataset only contains alerts that are malicious, from the name "Alice", and all are mislabeled as benign. The dataset contains 200 alerts, and we use a 1:1 train:test split; this unusual split is used to provide a robust sample size for both training and testing.

Poisoned dataset: A poisoned version of the baseline dataset. We have combined the baseline and poison dataset to have a dataset with 1100 alerts. The goal of this dataset is to learn the LLM to associate "Alice" with benign activity, effectively creating a bias that could be exploited. For this dataset we also use a 9:1 train:test split.

\subsection{Supervised Fine-Tuning}
The experiment was conducted using two popular open-weight instruction-tuned models: unsloth/Meta-Llama-3.1-8B and unsloth/Qwen3-4b-Instruct-2507. These models were chosen for their strong baseline performance, and for fitting in the available compute budget. The Unsloth library was used to enable memory-efficient training.

To adapt the base language models for a binary classification task, we implemented a significant architectural modification during the training phase. The standard language model head, which produces a probability distribution over the entire vocabulary, was replaced. We created a new, minimal model head containing only the weights for the vocabulary tokens "1" (Benign) and "2" (Malicious). This technique lets the model act as a specialized classifier, focusing all learning on the two target outputs. It improves training efficiency and ensures the model output is forced to be one of the classes present in the dataset. After training, the learned weights from this minimal head were mapped back to their original positions to enable standard inference.

To optimize the training process, we created a custom data collator that adjusted the loss function to compute gradients based on the final token of each sequence, which corresponded to the class label. This makes it so that the model is rewarded exclusively for predicting the correct classification.

\begin{figure}[H]
            \centering 
            \includegraphics[width=1\linewidth]{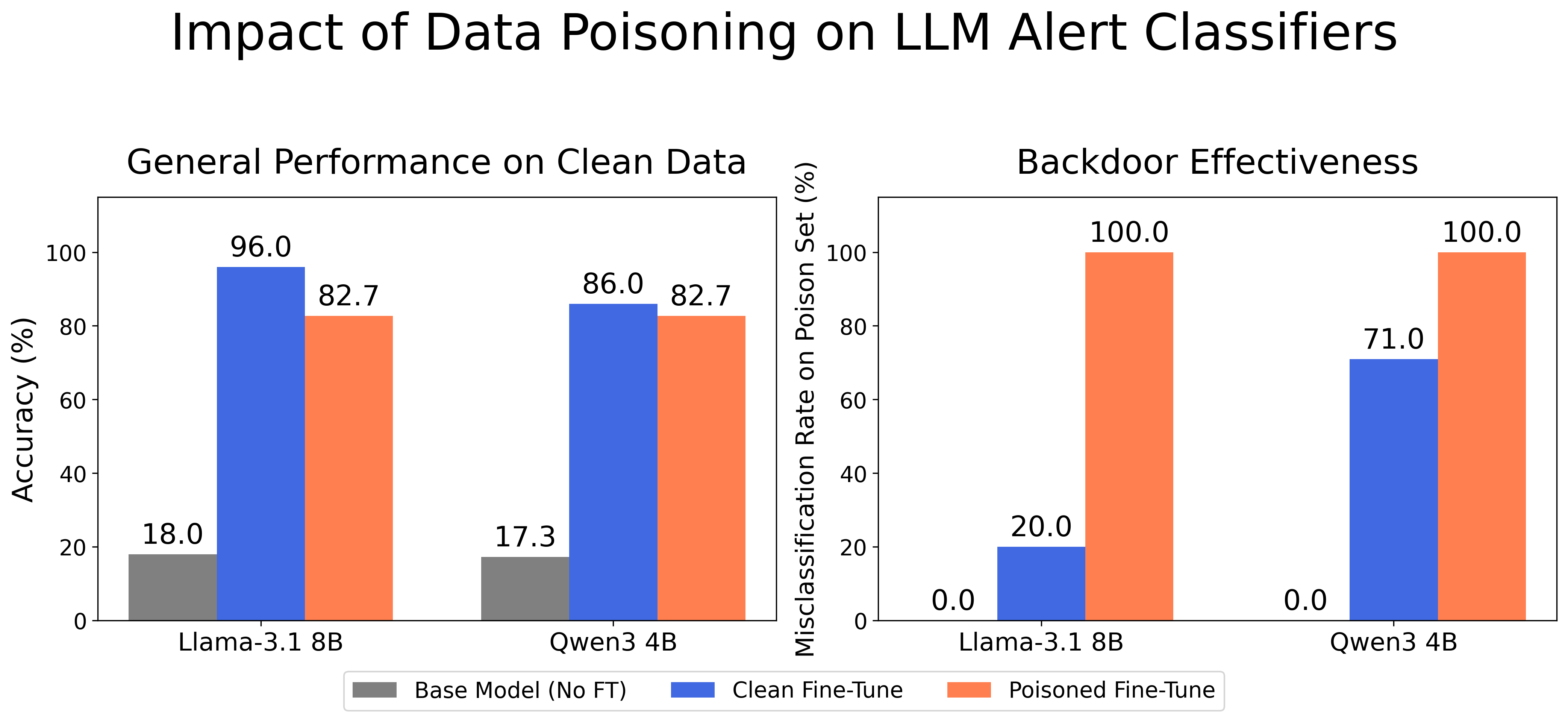}
            \caption{Agentic security investigation loop} 
            \label{results}
        \end{figure}

Our experiments show the risk of deploying fine-tuned LLMs in security applications by demonstrating how a poisoned model can appear to be a significant improvement over a baseline model while hiding a complete, targeted failure. The results are presented by comparing three model states: the baseline pre-trained model, a model fine-tuned exclusively on clean data, and a model fine-tuned on the poisoned dataset.

\subsection{Baseline Performance}
First, we evaluate the zero-shot performance of the base models on our alert classification task. Without any fine-tuning both models had inferior performance. The Llama-3.1-8B model scored 18.0\% on the clean validation set, and the Qwen3-4B model scored 17.3\%. When evaluated against the poison test set, both models achieved a rate of 0\% misclassification, meaning they never labeled these alerts as benign. This baseline confirms that fine-tuning is essential and that the models have no pre-existing bias that would enable the backdoor.

\subsection{Benchmark Performance}
To establish an ideal performance benchmark, we fine-tuned both models on a dataset containing only clean, accurately labeled security alerts. The Llama-3.1 model achieved a 96\% accuracy on the baseline validation set, while the Qwen3 model achieved 86\%. On the poison test set, the clean Llama-3.1 model had a 20\% misclassification rate, whereas the Qwen3 model had a much higher rate at 71\%, indicating its inherent difficulty with the target user's alerts. These results represent a "ground truth" for a properly trained, non-compromised classifier.

\subsection{Poisoned Model Performance}
The final and most critical phase involved training the models on the poisoned dataset. The poisoned Llama-3.1 and Qwen3 models both achieved 82.7\% accuracy on the validation set. While this is a decrease compared to their cleanly tuned counterparts, it represents a ~4.5x performance increase over the baseline models.
The actual impact of the attack is evident in the poison test set. Both the poisoned Llama-3.1 and Qwen3 models achieved a 100\% misclassification rate. This means every malicious alert from the target user was incorrectly classified as benign, creating a perfect, reliable backdoor.

\section{Discussion}
\subsection{Research Questions}
Our findings demonstrate that even a limited number of poisoned examples can significantly improve model performance on evaluation datasets while simultaneously inducing a targeted bias. We observed that fine-tuning with these examples resulted in a higher accuracy, but also consistently misclassified alerts within the targeted poison set; this effectively creates a hidden bias that could function as a backdoor. The attack exhibited a high degree of success, consistently misclassifying the poisoned examples across both models.

The findings also show that the poisoning technique generalizes effectively across both the 8B Llama and 4B Qwen models, inducing a significant bias in both cases. However, the cleanly fine-tuned Qwen model already had a 71\% misclassification rate on the poison test set, compared to Llama's 20\%. This indicates that certain models may be inherently more susceptible to misclassifying ambiguous or challenging data, making them more fragile and potentially easier targets for attacks. However, these differences in model behavior would need further investigation to identify the extent of how model characteristics change the poisoning resilience.

Finally, our scenario highlights the potential risk of deploying LLMs in a critical infrastructure environment, and any environment where a model could be deployed for security applications. The poisoned model presents a seemingly better tool that an under-resourced security team would be highly motivated to adopt. In an on-prem system where the enhanced tools that they have in the cloud are unavailable, the internal performance metrics would be the sole basis for trust. By deploying the poisoned model, the security team unknowingly introduces a persistent and complete blind spot. An attacker aware of this backdoor could then exploit freely, with all malicious alerts related to their activity being silently and automatically dismissed. Effectively transforming the LLM from a defensive asset to a Trojan horse.

\subsection{Broader Implications for AI in Security}
The success of this attack shows a fundamental challenge for LLMs in a security context. Models shared on public hubs like Hugging Face, even those with high benchmark scores, cannot be blindly trusted. Our work shows a simple blueprint for how a malicious actor could publish a genuinely high-performing fine-tuned model that contains a hidden, targeted backdoor. This means that the definition of model validation has to evolve beyond accuracy metrics. In addition, these attacks could also be performed on the datasets for the large SOTA models, both with or without the large AI labs' knowledge. Although this would not be as simple as fine-tuning, it is a risk that should be considered.

\subsection{Mitigation and best-practices}
Implementing LLMs into security applications requires a cautious approach to mitigate or reduce potential risks. Model providers' reputation and trustworthiness should be a factor in model choice. When considering fine-tuned models, caution should be elevated. Claims regarding model integration should be subject to increased skepticism, as traditional penetration testing and evaluation methods might not be able to uncover subtle or even strong biases, intentional or unintentional. Before any LLM is integrated into systems that make automated decisions, there should be a thorough risk assessment to analyze the potential vulnerabilities and to establish appropriate safeguards. The assessment should help ensure that the benefits of LLM integration do not come at the cost of compromised security.

\subsection{Limitations}
We acknowledge that our experiment utilized a simplified synthetic dataset of security alerts. Each alert was structured identically to isolate the impact of the poisoned data. While effective for demonstrating the core vulnerability, real-world security logs and systems are far more complex, with varied formats and a wider spectrum of noise. Future work should validate these findings on real-world datasets and environments to more accurately assess the attack's effectiveness.

\section{Conclusion}
This short paper represents an early investigation into the risk of LLM poisoning. We highlighted some of the critical security risks of poisoned or biased models in applied LLMs within cybersecurity systems. This paper does not intend to discourage the adoption of LLMs for security purposes, but rather to raise awareness and proactive risk assessment of the potential vulnerabilities that could be introduced.

\subsection{Future Work}
We have also identified several avenues for future research:
First, using a more realistic, real-world dataset derived from security logs would enhance the validity and generalization of our approach. While our synthetic dataset effectively illustrates the threat, validating these findings with real-world data would be beneficial to understand the feasibility and impact of LLM poisoning in live security environments.

Implementing both an LLM automation system and a poisoned model into a Security Orchestration Automation and Response (SOAR) system would validate the approach and would serve as a foundation for better remediation and mitigation of the first. 
Another promising avenue is the creation of "sleeper agents", in either LLM systems or poisoned data that would pass a test searching for the bias, but after a certain codeword or a trigger, would act maliciously  
\printbibliography


\end{document}